\documentclass[%
reprint,
showpacs,
showkeys,
amsmath,amssymb,aps,pra]{revtex4-1}
\usepackage{graphics}
\usepackage{epsfig}
\usepackage{dcolumn}
\usepackage{color}
\usepackage{bm}

\begin{document}

\preprint{APS/123-QED}

\title{Spin dynamics of Mn impurities and their bound acceptors in GaAs}

\author{M.R. Mahani, A. Pertsova and C.M. Canali}

\affiliation{Department of Physics and Electrical engineering, Linnaeus University, 391 82 Kalmar, Sweden}

\date{\today}

\begin{abstract}
We present results of tight-binding spin-dynamics simulations  
of individual and pairs of substitutional Mn impurities in GaAs. 
Our approach is based on the mixed quantum-classical 
 scheme for spin dynamics, with coupled equations of motions 
  for the quantum subsystem, representing the host, and 
 the localized spins of magnetic dopants, which are treated classically.   
 In the case of a single Mn impurity, we calculate explicitly the 
time evolution of the Mn spin and the spins of  
nearest-neighbors As atoms, where the acceptor (hole) state 
introduced by the Mn dopant resides. 
We relate the characteristic frequencies in 
the dynamical spectra to the two dominant energy scales of the system, 
namely the spin-orbit interaction strength and the value of the $p$-$d$ exchange 
coupling between the impurity spin and the host carriers. For a pair of Mn 
impurities, we find signatures of the indirect (carrier-mediated) 
 exchange interaction in the time evolution of the impurity spins. 
 Finally, we examine 
 temporal correlations between the two Mn spins and 
  their dependence on the exchange coupling and spin-orbit interaction strength, 
as well as on the initial spin-configuration and separation between the impurities. Our results 
  provide insight into the
dynamic interaction between localized magnetic impurities 
in a nano-scaled  magnetic-semiconductor sample, in the
extremely-dilute (solotronics) regime.
\end{abstract}


\maketitle


\section{\label{sec:level1}INTRODUCTION:\protect}

Recently, remarkable progress has been achieved in describing the electronic 
and magnetic properties of individual dopants in semiconductors, both 
experimentally\cite{yakunin_prl04,yazdani_nat06,bocq_prb13} and theoretically 
\cite{tangflatte_prl04,scm_MnGaAs_paper1_prb09,rm_cmc_d_2013}, offering exciting 
 prospects in future electronic devices. In view of novel potential 
applications, which involve communication between individual magnetic 
dopants, mediated by electronic degrees of freedom of the host, the focus 
of this research field has been shifting towards fundamental understanding 
and control of spin dynamics of these atomic-scale magnetic centers. 
Importantly, the development of advanced spectroscopic techniques has 
opened up the possibility to probe the dynamics of single spin impurities 
experimentally~\cite{KhajetooriansSci13}. Some specific examples include 
 inelastic tunneling spectroscopy of atomic-scale magnetic structures~\cite{Hirjibehedin19052006}, 
 optical manipulation of spin-centers in semiconductors~\cite{Fuchs11122009}, 
 magnetic resonance imaging~\cite{Rugar,Yacoby2,Yacoby3} 
 and 
 time-resolved scanning tunneling spectroscopy~\cite{Loth24092010}  
  on single spins. 
These advances pose new challenges for theory, calling 
for a fully microscopic time-dependent description of spin dynamics of 
individual spin impurities in the solid-state environment.\\
The most suitable approach to study the time evolution of spin systems, which is applicable 
 at ultrashort time ($\lesssim 100$~ps) and length ($\lesssim 1$~nm) scales and does not rely 
  on phenomenological parameters, is \textit{ab initio} 
 spin dynamics (SD). A natural framework for this approach is provided by the 
  extension of density functional theory (DFT) to time domain and noncollinear spins  
   (time-dependent spin DFT, or TD-SDFT), with several practical schemes 
   developed to date~\cite{PhysRevB.54.1019,PhysRevLett.75.729,
   PhysRevB.72.035308,Fahnle2005118,0953-8984-20-31-315203,PhysRevB.88.104423}. 
   However, due to computational demands, the system sizes that can be treated 
    with this approach are limited to only a handful of atoms~\cite{PhysRevB.88.104423}. 
   In practice, numerical implementations typically rely on approximations. 
   A well-known example is the adiabatic \textit{ab initio} SD model 
   by Antropov~\textit{et al.}~\cite{PhysRevB.54.1019,PhysRevLett.75.729}, 
    which is based on a Born-Oppenheimer(BO)-type approximation for spin degrees 
     of freedom in  materials with localized magnetic moments. 
   Because of the difference in the characteristic energy scales for itinerant 
    and localized spins, this model treats the directions 
    of the local magnetic moments as slow classical variables, while 
    averaging over the fast electronic degrees of freedom.  Such separation 
    leads to an equation of motion for the classical moments  
   interacting with an effective field.\\
    Antropov's SD model provides the theoretical framework for large-scale 
     implementations of atomistic spin dynamics, such as the one 
     by Skubic \textit{et al.}~\cite{0953-8984-20-31-315203}. 
      In the latter approach, the equation of motion for the localized 
       magnetic moments is augmented by a phenomenological 
     damping term (in analogy with Landau-Lifshitz-Gilbert equation) 
     and by a stochastic (Langeven) term, which accounts for 
     the effect of finite temperature. This scheme was used to investigate magnetic ordering and correlations 
      in  clusters of Mn-doped GaAs at finite temperature~\cite{PhysRevB.78.144419}. 
We also mention that, apart from standard atomistic spin-dynamics approaches, 
beyond mean-field dynamical models have been developed~\cite{1367-2630-11-7-073010}, which specifically address 
  ultrafast photoenhanced magnetization dynamics in dilute magnetic semiconductors (DMSs), 
  in particular (Ga, Mn)As~\cite{PhysRevLett.98.217401}.\\
In this work we employ the mixed quantum-classical SD model~\cite{StamenovaBook}, 
which is similar in spirit albeit different in some important aspects from 
Antropov's adiabatic SD, 
 to study the dynamics of individual and pairs of Mn impurities in GaAs. Although our model 
  assumes a partition into a quantum subsystem, representing the electrons of the host (GaAs), and 
  a classical subsystem, representing the localized magnetic moments of the dopants (Mn), the interaction 
   between the two subsystems is treated at the level of the Ehrenfest approximation~\cite{0953-8984-16-7-L03}, 
   as opposed to the BO approximation. This results in a system of coupled equations of motion, with the 
   two subsystems evolving simultaneously and  
   experiencing each other as time-varying classical fields. 
   As already known from its application to 
    electron-ion dynamics~~\cite{0953-8984-16-7-L03}, such scheme is able to capture non-adiabatic 
    effects on the fast electronic time scale (fs), 
   which are inaccessible by the BO approximation. Recently, Ehrenfest SD has been used 
    to study the effects of electrostatic gating in atomistic spin conductors~\cite{pertsovaPRB11}.\\
In order to describe the underlying electronic structure of the semiconductor 
 host, we use a microscopic tight-binding (TB) model for GaAs with parameters fitted to DFT calculations. 
 This gives our approach a computational edge compared to purely \textit{ab initio} SD, which relies  
  on the Kohn-Sham Hamiltonian~\cite{PhysRevB.54.1019}. The spins of Mn 
  dopants are introduced in the Hamiltonian as classical vectors 
 exchange coupled to the instantaneous spin-densities 
 of the nearest-neighbors As atoms, in the spirit of the $p$-$d$ exchange interaction. 
  In the static regime, this TB model has already proved successful in describing experimentally observed properties 
     of single Mn dopants and their associated acceptor states in GaAs~\cite{scm_MnGaAs_paper1_prb09,rm_cmc_d_2013}.\\
Here we explore the solotronics limit of DMSs~\cite{pm_nam11}
in the time domain by probing explicitly the  
  time evolution of individual substitutional Mn dopants in a finite nanometer-size cluster of GaAs. 
From the time-dependent spin-trajectories 
 of single-impurity spins, we 
 identify the characteristic energy scales of the dynamics, associated with intrinsic interactions 
present in the system, namely the $p$-$d$ exchange interaction and   
the spin-orbit interaction (SOI).
Furthermore, we study the time evolution of  two  Mn-impurity spins 
 upon an arbitrary perturbation of one of the spins. 
 The SD simulations allow us to address explicitly the time-dependent 
 (dynamic) indirect exchange interaction between the spins, 
which is expected to differ from its 
static counterpart~\cite{FranssonDynamic},  
and is relevant for nanostructures 
with magnetic impurities~\cite{CostaDynamic}. 
 We map out the temporal correlations 
   between the two localized spins, expressed in terms of a classical spin-spin correlation 
    function, as a function of the SOI, $p$-$d$ exchange interaction,  
   spatial separation, and the initial orientations of the  spins (
     e.g. ferromagnetic, antiferromagnetic and noncollinear configurations). 
   The variations in strength and time scale of the correlations, 
   inferred from behavior of the spin-spin correlation function for different system parameters, 
   can be used as an indicative measure  of dephasing of individual impurity spins due to their interaction with the 
   host and other impurities.\\
The paper is organized as follows. In Section II 
we describe the details of our theoretical approach. In Section 
III we discuss the results of  numerical simulations. 
 Section~\ref{1Mn} is concerned with the effect of SOI and exchange coupling 
  strength on the time evolution of a single Mn spin. 
  In Section~\ref{2Mn} we study the dynamical interplay 
  between two spatially-separated Mn impurities, coupled indirectly by  carrier-mediated exchange interaction. 
  Finally we draw some conclusions.\\

\section{THEORETICAL MODEL}
\label{theo_model}

\subsection{Tight-binding Hamiltonian}

We begin to lay out the computational framework of our SD simulations 
by first defining the respective Hamiltonians of the two subsystems. 
The time-dependent Hamiltonian of the GaAs host (quantum subsystem), 
 incorporating a number of substitutional Mn atoms on Ga sites, takes the following form

\begin{align}
\hat{H}_{el}(t) &  =\sum_{ii^{\prime},\mu\mu^{\prime},\sigma}t_{\mu\mu^{\prime}}^{ii^{\prime}}\hat{c}_{i\mu\sigma
}^{\dag}\hat{c}_{i^{\prime}\mu^{\prime}\sigma}+J_{pd}\sum_{m}\sum_{n[m]}\vec{S}_{m}(t)\cdot
\hat{\vec{s}}_{n}\nonumber\\
&  +\sum_{i,\mu\mu^{\prime},\sigma\sigma^{\prime}}\lambda_{i}\langle\mu
,\sigma|\hat{\vec{L}}\cdot\hat{\vec{s}}|\mu^{\prime},\sigma^{\prime}\rangle \hat{c}_{i\mu\sigma
}^{\dag}\hat{c}_{i\mu^{\prime}\sigma^{\prime}}\nonumber\\
&  +\frac{e^{2}}{4\pi\varepsilon_{0}\varepsilon_{r}}\sum_{m}\sum_{i\mu\sigma
}\frac{\hat{c}_{i\mu\sigma}^{\dag}\hat{c}_{i\mu\sigma}}{|\vec{r}_{i}\mathbf{-}\vec{R}%
_{m}|}+V_{\rm Corr},\label{class-ham}%
\end{align}
where $i$($i^{\prime}$) is the atomic index that runs over all atoms, $m$ runs over
the Mn atoms, and $n[m]$ over the nearest-neighbors (NN) of $m$-the Mn atom; $\mu$($\mu^{\prime}$) 
 labels atomic orbitals ($s$, $p_{x,y,z}$) and $\sigma$($\sigma^{\prime}$) is the spin index; 
 $t_{\mu\mu^{\prime}}^{ii^{\prime}}$ are the Slater-Koster parameters~\cite{chadi_prb77} and 
  $\hat{c}_{i\mu\sigma}^{\dag}$($\hat{c}_{i\mu\sigma}$) is the creation(annihilation) operator. 
Below, we briefly discuss the meaning of all the terms on the right-hand side of Eq.~(\ref{class-ham}). 
For a more detailed description the reader is referred to Ref.~\onlinecite{scm_MnGaAs_paper1_prb09}.

The first term in Eq.~(\ref{class-ham}) is the nearest-neighbors $sp^3$ Slater-Koster Hamiltonian~\cite{slaterkoster_pr54,papac_jpcm03} 
 that reproduces the band structure of bulk GaAs~\cite{chadi_prb77}.  
The second term represents the 
antiferromagnetic ($p$-$d$) exchange coupling between 
the Mn spin $\vec{S}_{m}$ (originating 
 from the $d$-levels of the dopant and treated here as 
a classical vector) and the nearest-neighbor  
As $p$-spins, $\hat{\vec{s}}_n =  1/2\sum_{\pi\sigma{\sigma'}} \hat{c}_{n\pi\sigma}^\dagger 
\vec{\tau}_{\sigma{\sigma'}} \hat{c}_{n\pi{\sigma'}}$, where  
$\vec{\tau}_{\sigma{\sigma'}}$ are elements of the Pauli matrices
 $\vec{\tau}=\{\tau^{\alpha}\}$ ($\alpha=x,y,z$) 
  and the orbital index $\pi$ runs over three As $p$-orbitals.
 We chose the value of the exchange coupling $J_{pd}=1.5$~eV, which has been 
 been reported in the literature~\cite{timmacd_prb05,ohno_sci98}. Note that 
  $\vec{S}_{m}$ is a unit vector and the magnitude of the Mn magnetic moment (5/2) is 
  absorbed by the exchange coupling parameter.
 
 The third term represents the one-body intra-atomic (on-site) SOI, where 
  $\hat{\vec{L}}$ is the orbital moment operator, $\hat{\vec{s}}$ is the spin operator,  
   $|\mu,\sigma\rangle$ are spin- and orbital-resolved 
    atomic orbitals, corresponding to atom $i$, 
 and $\lambda_i$ are the renormalized spin-orbit-splitting parameters~\cite{chadi_prb77} ($\lambda_{\mathrm{As}}=0.14$~eV, 
 $\lambda_{\mathrm{Ga}}=0.058$~eV and $\lambda_{\mathrm{Mn}}=
0.058/2$~eV). 

The fourth term represents the long-range repulsive Coulomb potential,  
 dielectrically screened by the host material, with 
 $\epsilon_r=12$ for bulk GaAs~\cite{Teichmann_prl08,Gupta_NanoLett}. $\vec{r}_i$ and 
  $\vec{R}_m$ denote the position of atom $i$ and the $m$-th classical spin, respectively. 
 The last term, $V_{\rm Corr}$, is the central-cell correction to the 
  impurity potential. This consists  
of the on-site part $V_{\mathrm{\rm on}}$, acting on the Mn ion, and 
 the off-site part $V_{\mathrm{\rm off}}$,    
  which affects the NN As atoms and is important for 
   capturing the physics of the $p$-$d$ hybridization, in addition to the exchange 
   interaction ($J_{pd}$). The value $V_{\mathrm{\rm on}}=1.0$~eV  
   is inferred from the Mn ionization energy, and we set $V_{\mathrm{\rm off}}=2.4$~eV 
    to reproduce the experimentally observed position 
     of the Mn-induced acceptor level in bulk GaAs~\cite{schairer_prb74,lee_ssc64,chapman_prl67,linnarsson_prb97}. 

We note that the time-dependence in the electronic Hamiltonian is carried by the Mn classical 
 spins, $\vec{S}_{m}(t)$. At time $t$, the classical Hamiltonian of the $m$-th Mn spin is written as

\begin{align}
H_{S}(t) &  =J_{pd}\sum_{n[m]}\langle \hat{\vec{s}}_{n}\rangle(t)\cdot
\vec{S}_{m}(t).
\label{ham}%
\end{align}
This describes the exchange coupling between $\vec{S}_{m}$ and the total instantaneous 
spin-density of the NN As $p$-spins, defined as a sum of expectation values  
$\langle\hat{\vec{s}}_n\rangle(t)=\mathrm{Tr}\left[\hat{\rho}(t)\hat{\vec{s}}_n\right]$, 
where $\hat{\rho}(t)$ is the density matrix of the electronic subsystem at time $t$. 



The dynamical properties of substitutional Mn atoms in a GaAs matrix are 
obtained by performing time-dependent SD simulations for 
a super-cell-type structure, consisting of a cubic cluster with $32$ atoms and 
 periodic boundary conditions applied in three dimensions. The equations of motion 
 which, together with the definitions of the quantum and classical Hamiltonians [Eqs.~(\ref{class-ham}) and (\ref{ham})], 
 make up the core of the SD simulation, are described in the next section.
 
\subsection{Equations of motion}

The time evolution of the quantum subsystem is governed by 
the Liouville-von Neumann equation for the density matrix, while the 
 localized spins, representing the Mn magnetic moments, evolve 
 according to its classical analogue. Thus the system of 
 coupled equations of motion reads

\begin{eqnarray}
\left\lbrace
 \begin{aligned}
\dfrac{d\hat{\rho}(t)}{dt} & =\dfrac{i}{\hbar}\left [\hat{\rho}(t),\hat{H}_{el}(t)\right ]\label{liouville}\quad\\
\dfrac{d\vec{S}_m(t)}{dt} & =\left\{ \vec{S}_m(t),H_S(t)\right\}\label{eqofmotion},
\end{aligned}
\right.
\end{eqnarray}
where $\left [\,,\,\right ]$ denotes the commutator and $\left\{\,,\,\right\}$ the Poisson bracket. Using the classical 
 analogue of the commutation  relations for $\vec{S}_m$~\cite{PhysRevA.22.1814}, we can calculate explicitly the Poisson bracket and 
 the classical equation of motion becomes
 
 \begin{align}
\dfrac{d\vec{S}_m(t)}{dt} & =\dfrac{J_{pd}}{S}\sum_{n[m]}\left\langle \hat{\vec{s}}_{n}\right\rangle(t)\times
 \vec{S}_{m}(t)
\label{eqofmotion1}%
\end{align}
where $S$ is the magnitude of  $\vec{S}_m$. 

As one can see from Eq.~(\ref{eqofmotion}), the quantum and the classical
 subsystems evolve according to their respective equations of motion but are coupled through 
 instantaneous exchange terms, which enter the time-dependent 
Hamiltonians $\hat{H}_{\rm el}(t)$ [Eq.~\ref{class-ham}] 
  and $H_S(t)$ [Eq.~\ref{ham}]. Such coupled system represents the Ehrenfest approximation to 
	spin dynamics~\cite{0953-8984-16-7-L03}. 
This is in contrast to the BO approximation, in which the fast (quantum) degrees of freedom are integrated out.
The equations of motion are integrated numerically using the fourth-order Runge-Kutta algorithm. The typical duration of the simulation 
 is of the order of $10$~ps. We chose a time step $dt=0.001$~fs, which insures that the total energy is 
 conserved within an error of $10^{-9}$~eV. 

\section{RESULTS AND DISCUSSION}

\subsection{Single Mn impurity}
\label{1Mn}

\begin{figure}
\centering
\includegraphics[scale=0.67]{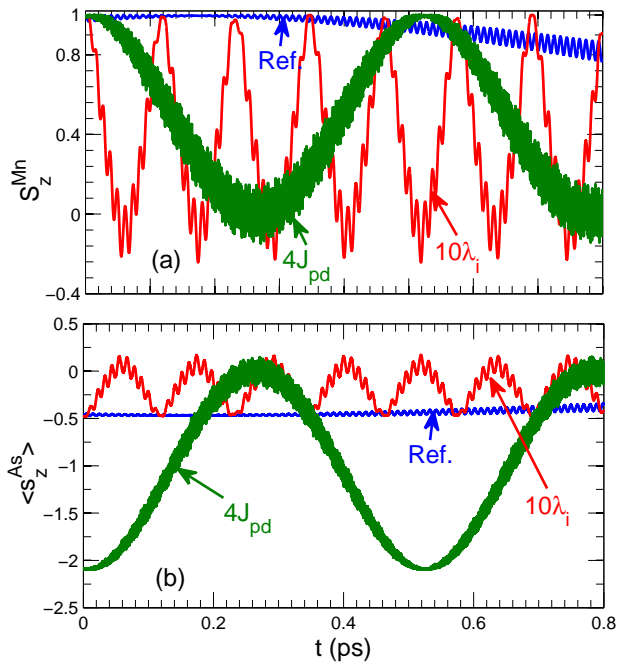}
\caption{(Color online) Time evolution of the $z$-components of the Mn 
spin (a) and the spin of the NN As atoms (b) for three different  
choices of exchange and SOI parameters: (\textit{i}) the reference case (see text for details), 
(\textit{ii}) the case in which $\lambda_{\mathrm{i}}$ is increased by a factor of $10$ and (\textit{iii})
 the case in which $J_{pd}$ is increased by a factor of  $4$. 
 Only the first 800~fs of the total simulation time of $10$~ps are shown.}
\label{fig:so-j-change}
\end{figure}

We first consider a single Mn impurity replacing a Ga in the center of a 4-atom GaAs cluster. 
 A smaller cluster size allows us to increase the simulation time up to $10$~ps, in order to understand the evolution of 
 the spins and all the energy scales involved in the dynamics. 
In Section~\ref{2Mn} the size of the cluster will be increased to 32 atoms to study the SD in the presence  of two Mn spins.\\ 
\begin{figure}
\centering
\includegraphics[scale=0.6]{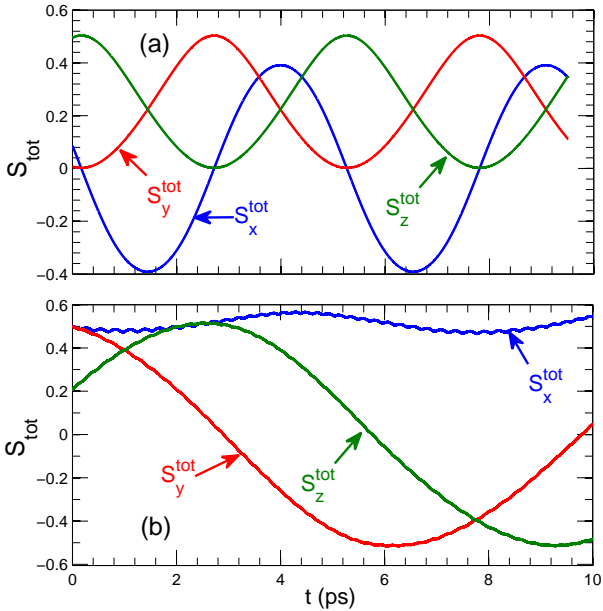}
\caption{(Color online) Time evolution of the total spin 
 of the system $\vec{S}^{\mathrm{tot}}$ for 
  two choices of the perturbation,  
	(a) $\theta=\phi=5^{\circ}$, 
and (b) $\theta=\phi=45^{\circ}$. 
Different components of the total spin are marked by arrows.}
\label{fig:tsp-ener} 
\end{figure}
\begin{figure}[htp]
\centering
\includegraphics[scale=0.54]{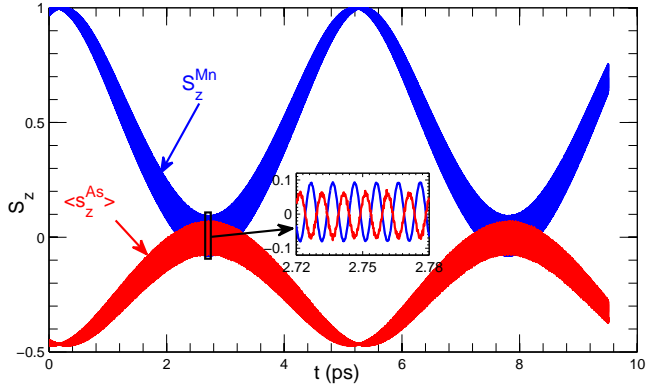}
\caption{(Color online) 
Time evolution of the $z$-components of the spins of the Mn impurity ($S^{\mathrm{Mn}}_z$) 
and its NN As atoms ($\langle s^{\mathrm{As}}_z\rangle$). The inset shows the zoom-in 
into smaller time scale.}
\label{fig:4at-Mn-As2} 
\end{figure}
Before the start of the simulation (at $t$=$0$), the orientation of the classical Mn spin 
  is fixed along the [001] direction. This corresponds to the $z$-axis, $\vec{S}_{m}(0)\parallel z$, or  
   $\theta=\varphi=0$ in spherical coordinates, where $\phi$ is the azimuthal angle and $\theta$ is the polar angle.   
  The calculations of magnetic anisotropy landscape for a Mn impurity in bulk GaAs,  described 
  with the classical-spin model used in this work, have shown that 
   the [001] direction is the easy axis, while the plane perpendicular to it ($x$-$y$) is the 
	hard plane~\cite{scm_MnGaAs_paper1_prb09}.  
  For this equilibrium orientation, the 
  electronic density matrix is constructed as $\hat{\rho}(0)=\sum_{\nu}f_{\nu}|\psi_{\nu}\rangle\langle\psi_{\nu}|$, 
  where $\{\psi_{\nu}\}$ are the eigenfunctions of $\hat{H}_{el}(0)$ (see Eq.~{\ref{class-ham}}) and $f_{\nu}$ 
  are Fermi-Dirac occupation numbers. The initial 
 NN As spin-density, entering the classical Hamiltonian in Eq.~(\ref{ham}), is calculated as 
 $\langle \vec{s}^{\mathrm{As}} \rangle(0)=\sum_{n[m]}\langle 
 \hat{\vec{s}}_n(0)\rangle=\sum_{n[m]}\mathrm{Tr}\left [ \hat{\rho}(0)\hat{\vec{s}}_{n} \right ]$.\\
In order to  initiate the SD, the Mn spin is tilted from its preferential axis by angles 
$\theta$ and $\varphi$. This procedure represents an arbitrary external 
perturbation, applied locally to the Mn magnetic moment, e.g. a 
laser pulse  or an external magnetic field. 
As an output of the simulation we obtain the 
time-dependent Cartesian components of the Mn spin, which will be 
denoted as $\vec{S}^{\mathrm{Mn}}$ throughout this section, as well as 
 the time-dependent expectation value of the spin at any given 
atom $i$ of the cluster, $\langle\vec{s}_i\rangle$. 
We will focus in particular on the 
 total spin $\langle \vec{s}^{\mathrm{As}}\rangle$ of the NN As atoms, 
where the Mn-induced spin-polarized acceptor state resides, and 
the total spin of the system defined as $\vec{S}^{\mathrm{tot}}=\vec{S}^{\mathrm{Mn}}+
\sum_i\langle\vec{s}_i\rangle$, where $i$ runs over all Ga and As atoms.\\ 
Figure~\ref{fig:tsp-ener} shows the time evolution of the three components of the total spin 
$\vec{S}^{\mathrm{tot}}$ 
 for two different choices of the perturbation, namely $\theta=
\varphi=5^{\circ}$ and $\theta=
\varphi=45^{\circ}$. The first choice corresponds to 
a weak perturbation since at the start of the simulation the Mn spin deviates 
only slightly from its preferential axis. We will use this 
type of perturbation in this section.
Throughout the next section,~\ref{2Mn}, we will use a slightly stronger
perturbation, $\theta=\varphi=10^{\circ}$, unless specified otherwise.\\
As one can see from Fig.~\ref{fig:tsp-ener}(a), all three components 
of $\vec{S}^{\mathrm{tot}}$ exhibit long-period ($\approx 5.5$~ps) oscillations, 
which appear by turning on the SOI. Note that without SOI 
the three components of the total spin are constants of motion and do 
not change during the time evolution.  
We conclude that in the case of a weak perturbation the 
dynamics of the total magnetic moments is mainly driven by the SOI. 
 This is expected since the Mn spin remains in equilibrium 
 with the spins of the NN As atoms.     
 However, a strong perturbation ($\theta=\phi=45^{\circ}$) 
 brings about the interplay between $\vec{S}^{\mathrm{Mn}}$ 
and $\langle \vec{s}^{\mathrm{As}}\rangle$, governed by the 
 exchange coupling $J_{pd}$. This results 
 in short-period ($\approx 500$~fs) oscillations, superimposed on the long-period  
and large-amplitude precession due to SOI [see Fig.~\ref{fig:tsp-ener}(b)].\\
%
A similar effect, namely the appearance of pronounced 
oscillations due to $J_{pd}$, will be observed if we artificially scale up the exchange constant. 
Note also that the strong perturbation forces the $z$-component 
of the total spin to oscillate between two easy axes 
(parallel or anti-parallel to the $z$-axis), while in the case 
of the weak perturbation $\vec{S}^{\mathrm{tot}}$ 
remains above the $x$-$y$ plane ($S^{\mathrm{tot}}_z>0$).\\
The comparison between panels~\ref{fig:tsp-ener}(a) and (b) also indicates that the 
resulting dynamics \  
and the oscillation frequency 
are sensitive to initial conditions, especially for 
strong perturbations beyond the linear-response regime.\\
Figure~\ref{fig:4at-Mn-As2} shows the time evolution of the $z$-components of the 
Mn spin and the spin of the NN As atoms. Similarly 
to $S^{\mathrm{tot}}_z$ [Fig.~\ref{fig:tsp-ener}(a)], 
 the dynamics is governed by the long-period precession.   
 However, the zoom-in into smaller times reveals fast 
oscillations due to $J_{pd}$. $S^{\mathrm{Mn}}_z$ 
and $\langle s^{\mathrm{As}}_z\rangle$ are oscillating 
 in anti-phase, in accordance with 
the antiferromagnetic nature of the $p$-$d$ exchange coupling. \\
In order to understand how the characteristic energy scales of the system 
 control its dynamics, 
 we perform SD simulations with modified values of exchange and SOI 
parameters. 
As a references set of parameters, we consider 
the values of the spin-orbit splittings $\lambda_i$ 
 and the exchange interaction $J_{pd}$ typically used in our TB model for (Ga,Mn)As. 
Next, we consider two cases: 
(\textit{i}) the exchange interaction is unchanged and the SOI strength is $10\lambda_i$ and 
(\textit{ii}) the SOI is unchanged and the exchange interaction is $4J_{pd}$. 
The time evolution of the Mn and NN As spins, calculated with the 
 reference set and with the modified parameters,    
is shown in Figs.~\ref{fig:so-j-change}(a) and (b), respectively.\\ 
\begin{figure}
\centering
\includegraphics[scale=0.55]{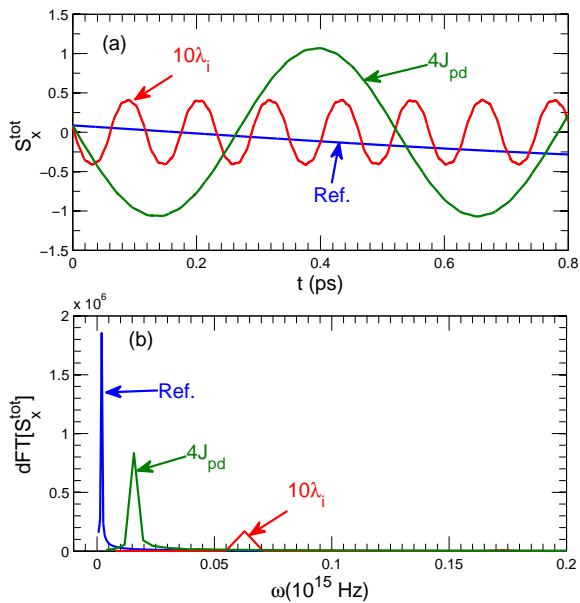}
\caption{(Color online) (a) Time evolution of the $x$-component of the total spin of the 
system and (b) its discrete Fourier transform,     
for three different choices of exchange and SOI parameters.  
 Notations are the same as in Fig.~\ref{fig:so-j-change}.}
\label{fig:energy}
\end{figure}
From the simulation with the reference set, we identify two main periods (frequencies), 
namely the long-period (low-frequency) 
oscillations and the short-period (high-frequency) oscillations. 
 As mentioned before, these two characteristic periods are most likely associated with 
 SOI and with exchange interaction, respectively, since $J_{pd}$ is several orders 
of magnitude larger than $\lambda_i$. 
This is further confirmed by the simulations with the modified parameters. 
 Increasing the SOI strength, while keeping $J_{pd}$ unchanged, leads to a decrease of 
the long period. At the same time 
 the period of rapid oscillations, superimposed 
 on the dominant long-period precession, remains practically unchanged.\\
Increasing the exchange parameter by 
 a factor of $4$ yields a significant decrease of the short period oscillations. 
However, there is also a noticeable change (decrease) in the long period oscillations. 
This is 
due to the fact that, in principle, one should not expect the two energy scales 
to affect the dynamics in a completely independent way. 
 The complex dynamics of our combined quantum-classical 
system results from the interplay 
of the interactions (rather than simply from   
a superposition of harmonic motions with different frequencies). 
 The change in $J_{pd}$ has the strongest effect on both 
short-period and long-period dynamics since it is the largest energy scale 
 in the system.

The effect of the two energy scales can be also observed in the 
time evolution of the total spin. The time-dependent $x$-component 
of the total spin and its discrete Fourier  transform (dFT) are shown in Fig.~\ref{fig:energy}(a) and (b), respectively. For all three 
choices of parameters the long-period oscillations due to SOI are the most pronounced. 
 The period of these oscillations is approximately $5.5$~ps for the reference set, which is equivalent to the frequency of $10^{12}Hz$. The latter can be associated with 
the position of the major peak in the corresponding dFT [see Fig.~\ref{fig:tsp-ener}(b)]. 
Increasing the SOI strength leads to an increase of the characteristic period 
and to a shift of the characteristic frequency to the higher range. 
An increase of the exchange coupling  parameter has a similar effect although the shift is smaller.

\subsection{Two Mn impurities}
\label{2Mn}

\begin{figure}[htp]
\centering
\includegraphics[scale=0.3]{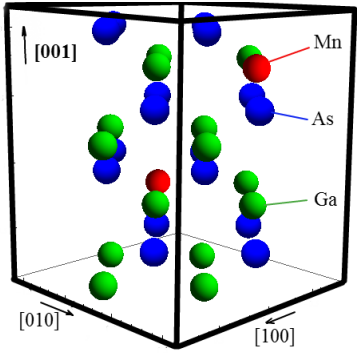}
\newline
\caption{(Color online) A finite cluster of GaAs, consisting of $32$ atoms, with two Mn impurities. 
 The separation between the impurities is $d\approx 0.8$~nm.}
\label{fig:structure} 
\end{figure}

In this section we focus on the dynamical interaction between 
 two spatially separated Mn spins. We use a GaAs cluster illustrated in Fig.~\ref{fig:structure}. 
The effects of the interaction parameters (exchange coupling and SOI), initial configuration 
 of the spins and the separation between the impurity atoms on the dynamics will be investigated.
 Throughout this section the two classical spins are labeled as $\vec{S}^1$ and $\vec{S}^2$.

\begin{figure}
\centering
\includegraphics[scale=0.5]{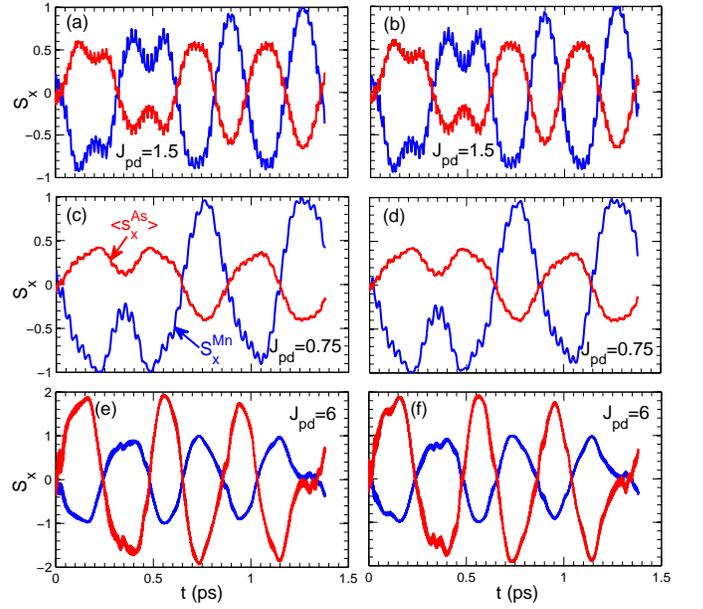}
\caption{(Color online) The time evolution of the $x$-components of the two Mn spins (blue curves) and the spins of their respective NN As atoms (red curves) for (a,b) $J_{pd}=1.5$~eV, (c,d) $J_{pd}=0.75$~eV and 
(e,f) $J_{pd}=6$~eV. Left panels are for the first (tilted) Mn spin and 
 right panels are for the second Mn.}
\label{fig:2MnAS_jx} 
\end{figure}

We first consider the situation when the two 
Mn spins are initially pointing along the $z$-direction. At the start of the simulation, one of 
the Mn spins is tilted by a small angle, $\theta=\varphi=10^{\circ}$, from its initial 
orientation. Figure~\ref{fig:2MnAS_jx} shows the time evolution of the 
$x$-components of both Mn spins and the spins of the corresponding NN As atoms for three 
different values of 
the exchange interaction (similar curves are obtained 
 for other spin-components). The dynamics is almost identical for the two Mn spins, 
which is a signature of the ferromagnetic coupling between the two.  
Each of the localized spins is coupled antiferromagnetically to 
 the spins of its NN As atoms, resulting in characteristic 
anti-phase oscillations.
  
\begin{figure}
\centering
\includegraphics[scale=0.55]{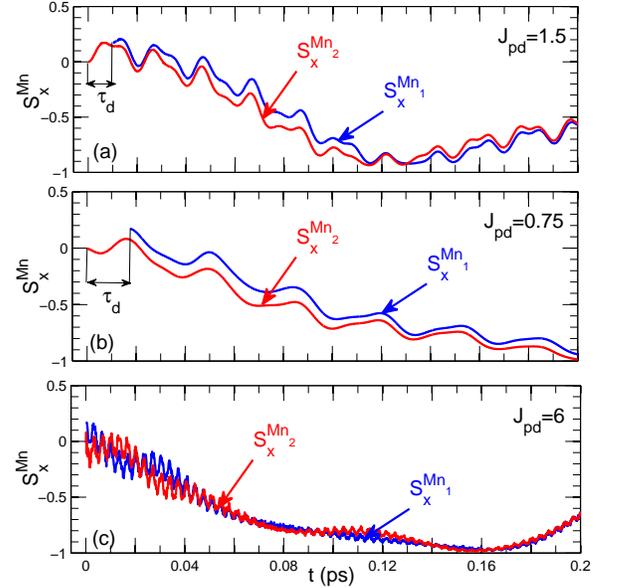}
\caption{(Color online) Time evolution of the $x$-components of the two Mn 
spins during the first 200~fs from the start of the simulation, for (a) $J_{pd}=1.5$~eV, (b) $J_{pd}=0.75$~eV and (c) $J_{pd}=6$~eV. Notations are the 
same as in Fig.~\ref{fig:2MnAS_jx}. Note that the blue curves (tilted spin) in panels (a) and (b) are shifted by 
the time delay $\tau_{\mathrm{d}}$ 
(see text for details) to show 
the similarity in the time evolution of the two spins at later times.
}
\label{fig:2Mn_jx_zoom} 
\end{figure}

 The SD simulations reveal a short delay, or response time (of the order of $1$~fs for $J_{pd}=1.5$~eV), between 
the perturbation and the response of the second spin. The response time decreases with increasing the 
 strength of the exchange interaction $J_{pd}$ between each of the individual Mn 
spins and the spins of the NN As atoms.  
This means that the spin perturbation, generated by the precession of the first (tilted) spin is transferred  more efficiently 
 to the second spin, when the exchange interaction between the Mn impurity spins and the host carriers is  stronger. 
 As a result, the effective dynamical interaction between the two impurity spins 
is enhanced. We can identify a characteristic time-interval $\tau_d$ (time delay) such that 
 when the two curves, corresponding to the two Mn spins, are shifted by $\tau_d$, they appear to oscillate in a similar way 
 ($\tau_d=10$~fs for $J_{pd}=1.5$~eV). This is essentially the time needed to establish the correlation between the two Mn spins. The value of $\tau_d$ 
 decreases with $J_{pd}$. When the time delay is taken into account, the short period oscillations of the spins can be simply superimposed on each other, as shown in 
 Figs.~\ref{fig:2Mn_jx_zoom}(a) and (b).

%
%
In order to understand how the electronic subsystem (GaAs host) mediates 
the propagation of spin disturbance between the localized spins, depending on the system parameters, 
 we calculate the classical spin-spin correlation function~\cite{StamenovaBook}\\
\begin{align}
C(S^1_x,S^2_x,\Delta{t}) &  = \frac{1}{N}\int{S^1_x(t)S^2_x(t+\Delta{t})dt}\;,%
\label{corr}
\end{align}
where $\Delta t$ is the \textit{time lag} and $N$ is  the normalization factor
\begin{align}
N &  = \left\{ \int{\left[S^1_x(t)\right]^2dt}\right\}^{\frac{1}{2}}\cdot\left\{\int{\left[S^2_x(t)\right]^2dt}\right\}^{\frac{1}{2}}.%
\label{norm}
\end{align}
Here we focus on the temporal correlations between the transverse ($x$) components of the spins, corresponding 
to deflections away from the easy axis. However, longitudinal components have also been considered 
 and the resulting correlation functions show a similar behavior.
 
A few remarks are in order about the definition and the meaning of the classical cross-correlation 
function in Eqs.~(\ref{corr}) and (\ref{norm}). 
In principle, for continuous functions of time, the integrals in 
 the cross-correlation function are taken over the total simulation time $T$ in 
 the limit 
$T\rightarrow\infty$, i.e. $\int \rightarrow \lim\limits_{T\rightarrow\infty}\int\limits_{0}^{T}$. 
 For functions sampled on a finite time-interval $T$, the integrals are replaced by 
finite sums and the correlation function becomes~\cite{Matlab} 
\begin{align}
C(S^1_x,S^2_x,\Delta{t}) &  = \frac{1}{N}\sum_{i=0}^{N_{st}-\Delta{t}}{S^1_x(t_i)\cdot S^2_x(t_i+\Delta{t})}\;,%
\label{corr1}
\end{align}
where $t_i=i\cdot dt$, $\Delta t=j\cdot dt$ and $j$ runs from $1$ to $N_{st}$, with $N_{st}=T/dt$ being the total number of time steps. The normalization factor now reads 
\begin{align}
N &  = \left\{\sum_{i=0}^{N_{st}}{S^1_x(t_i)\cdot S^1_x(t_i)}\right\}^{\frac{1}{2}}\cdot
\left\{\sum_{i=0}^{N_{st}}{S^2_x(t_i)\cdot S^2_x(t_i)}\right\}^{\frac{1}{2}}.%
\label{norm1}
\end{align}
%
%
In this work we adopt the latter definition.

The cross-correlation function $C$, defined above, measures the similarity in the temporal 
 profiles of two signals (in this case the two Mn spins) over the total simulation time, 
 as a function of the time lag $\Delta{t}$ applied to one of them (this simply means that one of the 
 signals is probed at a later time $t+\Delta{t}$ 
 compared to the other one). The properties of the correlation function are described below.
 
 The amplitude of $C$ at a given $\Delta t$ is a direct measure of the correlation between 
 the two signals: if the correlation function has a node ($C=0$), the two signals are completely uncorrelated, 
 while $C=1$ means that the signals are essentially identical; note that due to the normalization factor, 
 $\mathrm{max}\left\{|C|\right\}\le 1$.
 Let us first consider two sinusoidal signals, which have the same period but 
  differ by a phase. The correlation function $C$ in this case is also a sinusoidal function with the same period.  
 The value of $C$ at zero time lag depends on the 
phase difference between the two signals and varies in the interval $[-1,1]$. 
 In the ideal case of an infinite integration interval, such correlation function 
  would oscillate forever without decay, with an amplitude changing between $-1$ to $1$. 
  If we now add a random contribution to the second signal, the amplitude of oscillations 
   of the correlation function will decrease, since the coherence between the two signals 
   is hindered.
   
 However, for signals sampled on a finite time-interval,
   an additional issue affects the amplitude of the 
   correlation function. Indeed,  
 the integration interval, i.e. the number of terms in the finite sum in 
 Eq.~(\ref{corr1}), becomes smaller as the time lag increases. Therefore, the amplitude of 
the correlation function decreases with the time lag 
and vanishes for $\Delta t$ equal to the total simulation time 
(even for two identical sinusoidal signals). 
This decay is linear as a function of the time lag in the above example  
and it is negligible when the maximum time lag is much smaller than the integration interval.\\
\begin{figure}
\centering
\includegraphics[scale=0.72]{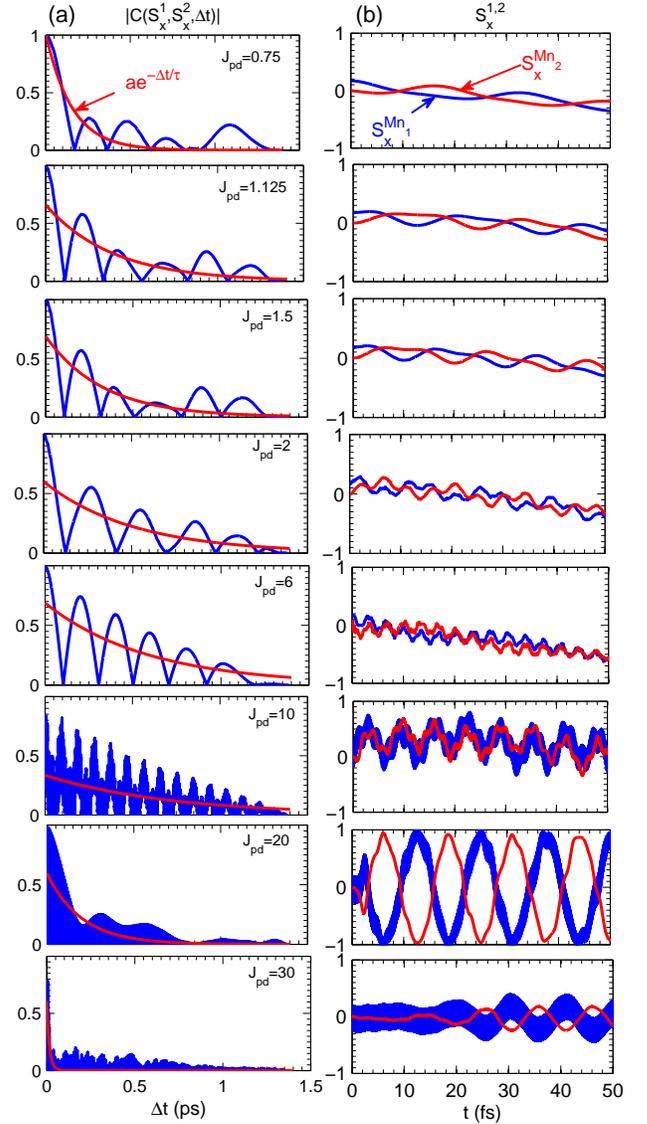}
\caption{(Color online) (a) The absolute value of 
the correlation function $\left|C(S^1_x,S^2_x,\Delta{t})\right|$, 
 plotted as a function of the  
time lag $\Delta t$, for different values of $J$. 
 The fitting exponential functions $a\cdot e^{-\Delta{t}/\tau}$ 
 are shown in red. 
 (b) Time evolution of the $x$-component of the first (blue curve) 
 and second spin (red curve).}
\label{fig:2Mn_corr} 
\end{figure}
Hence, we conclude that for realistic signals, two distinct factors affect 
the amplitude of the correlation function: 
(\textit{i}) the coherence, or the phase difference between the two signals at a given time lag, 
which may result in an increase or a decrease of the amplitude and is physically meaningful, and (\textit{ii}) 
the decay of the amplitude as a function of the time lag due to finite integration interval, which is inherent in   
the definition of the correlation function for sampled signals [see Eq.~(\ref{corr1})].
In practice, it is problematic to separate the artificial decay due to finite integration intervals 
 from the decay due to the gradual loss of coherence (dephasing) between the signals with increasing time lag. However, 
  any increase or non-linear decay of $C$ can be associated with the level of coherence. 
  It is also meaningful to compare the oscillating behavior and the decay of the correlation function 
  for different system parameters, with respect to a reference parameter set. As shown below, the information extracted 
  from the correlation function will be used mainly as a comparative measure of the temporal 
  correlations between the two Mn spins.\\
%
Figure~\ref{fig:2Mn_corr}(a) shows the absolute value of 
 $C(S^1_x,S^2_x,\Delta{t})$ as a function of 
the time lag $\Delta t$, for different 
values of the exchange coupling $J_{pd}$. The 
curves are fitted to an 
exponential function 
\begin{align}
f(\Delta t) &  = a\cdot e^{-\Delta{t}/\tau}.%
\label{fitt}
\end{align}
where $a$ is a constant and $\tau$ is the 
 characteristic decay time, with parameters listed in Table~\ref{tab:fitted}. 
The constant $a$ 
 indicates the degree of 
correlation between the two spins while $\tau$ 
is the combined measure of the intrinsic decay due to limited integration interval and
the fluctuations due to the phase difference between the two spins for increasing time lag. 
 While the period of the oscillations and the constant $a$ are more robust,
the value of $\tau$ depends on the integration interval $T$ (here $T=2$~ps) and it gradually goes to zero
as the time lag becomes comparable to $T$. 
Although the value of $\tau$ for a given $J_{pd}$ does not necessarily indicate the decay 
due to dephasing of the two spins, comparing this value for different panels in Fig.~\ref{fig:2Mn_corr}(a)  
 provides insight into how quickly the correlation dies out for different $J_{pd}$.\\
\begin{figure}
\centering
\includegraphics[scale=0.5]{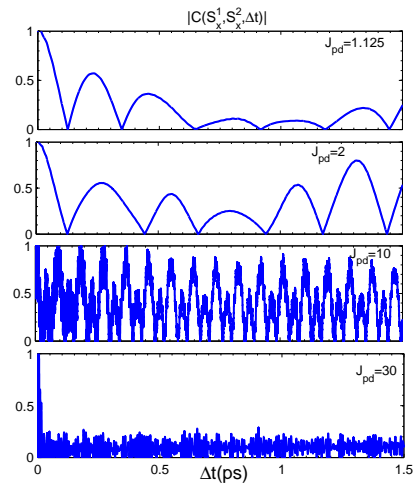}
\caption{(Color online) (a) The absolute value of 
the correlation function $\left|C(S^1_x,S^2_x,\Delta{t})\right|$, 
 plotted as a function of the  
time lag $\Delta t$, for different values of $J_{pd}$. 
Note that the integration scheme for the correlation function used  
 here differs from the rest of the paper (see text for details).}
\label{fig:corr_comp} 
\end{figure}
\begin{table}
\caption{\label{tab:fitted}Parameters for the fitting functions  
in Fig.~\ref{fig:2Mn_corr}(a). The correlation function for different 
$J_{pd}$.}
\begin{ruledtabular}
\begin{tabular}{ccc}
$J_{pd}$~(eV) & a & $\tau$~(fs) \\
\hline
0.75    &1.0    &155 \\
1.125   &0.66   &389 \\
1.5     &0.69   &328   \\
2.0     &0.6    &508   \\
6.0     &0.69   &586   \\
10.0    &0.34   &718   \\
20.0    &0.6    &200   \\
30.0    &0.61   &15   \\
\end{tabular}
\end{ruledtabular}
\end{table}
To further clarify the issues related to the definition of the correlation 
 function, we present in Fig.~\ref{fig:corr_comp}  the absolute value of 
 $C(S^1_x,S^2_x,\Delta{t})$ calculated using a different integration scheme. 
  Here the total simulation time is $T=4$~ps and the integration interval 
  is $[0+\Delta t,T/2+\Delta t]$~ps, i.e. the sum in Eq.~(\ref{corr1}) runs 
  from $i=j$ to $i=N_{st}/2+j$. 
Therefore, as the time lag increase, the integration interval 
remains constant ($T/2=2$ ps). The results obtained with this 
integration scheme clearly show that  the 
decay of the correlation function seen in Fig.~\ref{fig:2Mn_corr}(a) is, indeed, 
partly caused by the dephasing between the two spins at larger time lags. 
Notably, the decay in the first two panels of Fig.~\ref{fig:corr_comp}, which is \textit{not} caused by the decrease 
in the integration interval, 
 is in agreement with the corresponding panels in Fig.~\ref{fig:2Mn_corr}(a).
The slower decay for $J_{pd}=10$~eV and the very fast decay for $J_{pd}=30$~eV are also in agreement with Fig.~\ref{fig:2Mn_corr}(a) 
 and with the values of the fitting parameters in Table~\ref{tab:fitted}. 
We should also mention that for all cases considered, the exponential fitting curve  
 has better fitting statistics than a linear fit. This further confirms that $\tau$ captures part of  
 the decay coming from the dephasing of the two spins. 
 However, in the rest of the paper, we use the value of $\tau$ as a comparative rather than absolute measure of the dephasing.\\
%
We will now discuss in detail the behavior of the correlation function 
 for different values of $J_{pd}$, shown in  Fig.~\ref{fig:2Mn_corr}(a).   
 The peak in the amplitude of the correlation function occurs  
at a very early time compared to the total simulation time. 
 This confirms the observation that the  
  transfer of the spin perturbation between the two spins, 
  mediated by the host carriers, happens on a very fast 
	 time scale, e.g. $\sim 10$~fs 
 (see Fig.~\ref{fig:2Mn_jx_zoom}). 
The degree of correlation at $\Delta t=0$, $a$, 
is maximum for 
 $J_{pd}=0.75$ and nearly constant ($a\approx 0.6-0.7$) for other values (except 
  a somewhat special intermediate case of $J_{pd}=10$~eV). In contrast, 
	the decay time $\tau$ varies significantly with $J_{pd}$. It reaches 
	 its maximum for $J_{pd}\approx 10$~eV and decreases afterward. 
	 For exchange interactions in this range, 
the system is in a transient regime,  
before it undergoes a transition to a new dynamical state in which the Mn spins are effectively decoupled.
	 
The long-period oscillations of the correlation function for $J_{pd}<10$~eV indicate  
that the frequency of precession of the tilted spin of precession 
of the tilted spin is not comparable to, 
i.e. it is much smaller   
 than the characteristic electronic frequency or ``electron ticking time''. The latter is  
  the inverse of the time 
	 needed for the spin perturbation to travel forth 
	and back between the two spins.  
	For $J_{pd}\gtrsim 10$~eV these frequencies become comparable, and the correlation 
function oscillates very rapidly as a function of the time lag. It is 
 also for a $J_{pd}$ around this value that   
   the temporal correlations become long-ranged (the value $\tau$ is maximum in Table~\ref{tab:fitted}). 
	This essentially means that the
	 dynamical 
	coupling between the localized spins is the strongest in 
	this regime. 
Going over to yet larger values of $J_{pd}$, the carriers are too localized 
to transfer the spin perturbation efficiently, and therefore
 the spins become decoupled. This 
 is also consistent with the 
 short decay time of the correlation 
 function $\tau\approx 15$~fs for $J_{pd} = 30 $ eV. 

\begin{figure}
\begin{minipage}[h]{0.49\linewidth}
\centering
\includegraphics[scale=0.57]{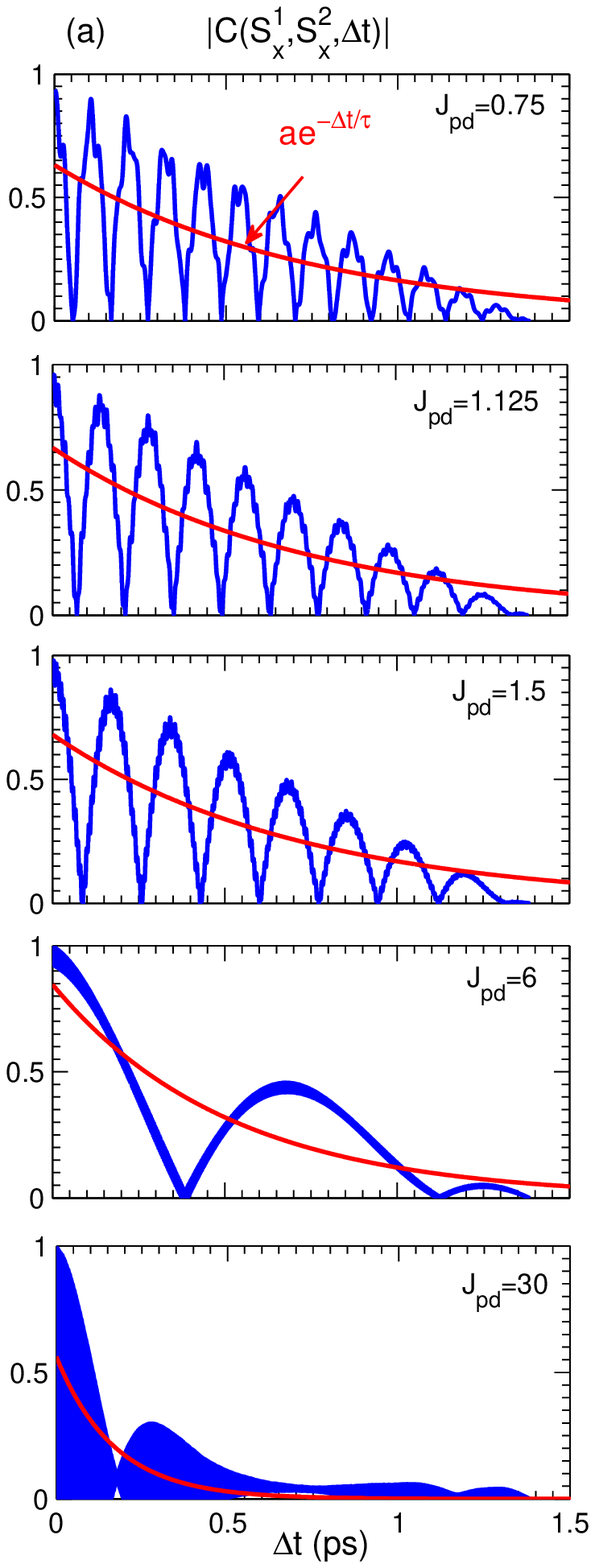}
\end{minipage}
\begin{minipage}[h]{0.49\linewidth}
\centering
\includegraphics[scale=0.57]{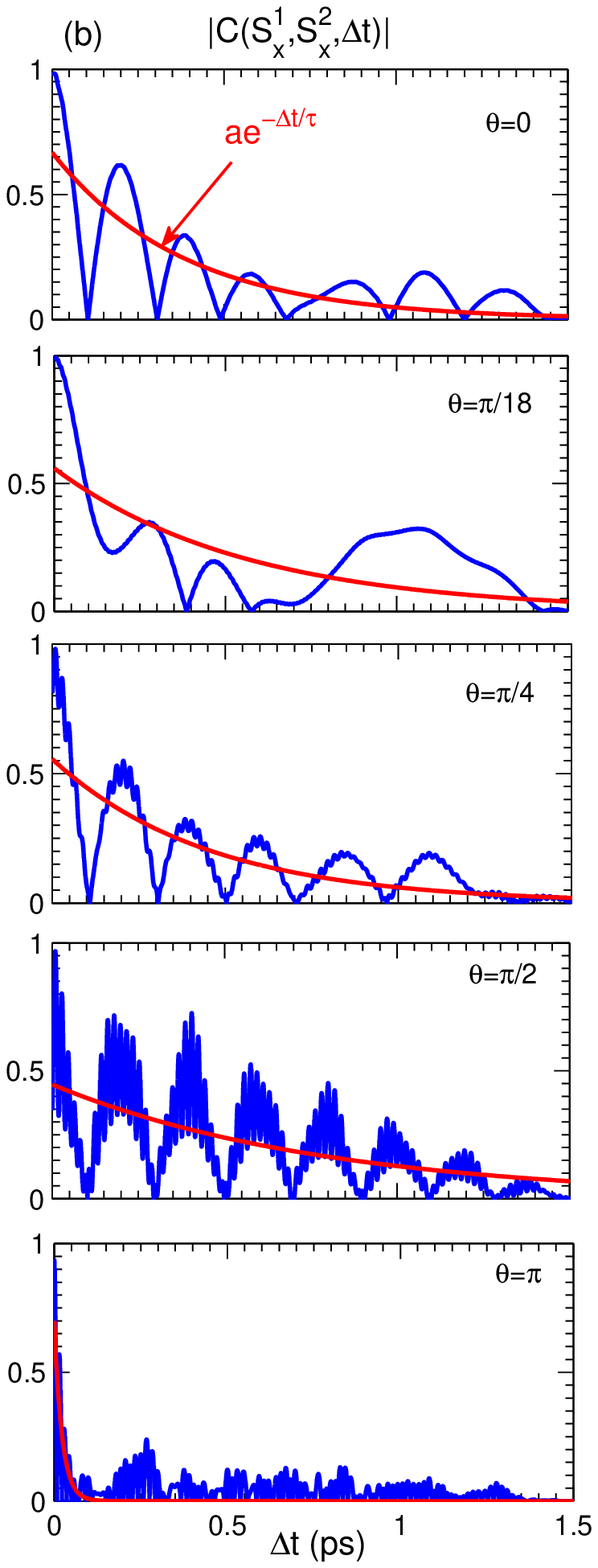}
\end{minipage}
\caption{(Color online) (a) The absolute value of 
the correlation function $\left|C(S^1_x,S^2_x,\Delta{t})\right|$, plotted as a function of the  
time lag (blue curves), and the corresponding 
 exponential fitting functions (red curves),  
 for different values of $J_{pd}$. The Mn spins are nearest-neighbors. 
 (b) The same correlation function for the 
case when the second spin is initiated at different angles ($\theta$ varies from $0$ to $\pi$ and $\phi=0$, 
$J_{pd}$ is fixed at the reference value).  
}
\label{fig:2Mn_ang} 
\end{figure}

This behavior is consistent with time evolution of the transverse components   
of the spins [right panels of Fig.~\ref{fig:2Mn_corr}]. 
 Except for a short time delay for small values of $J_{pd}$,  
  the motion of the two spins is correlated for $J_{pd}\lesssim 10$~eV.  
 However, for very large values of $J_{pd}$ the second spin completely looses 
 the high-frequency component and even oscillates in anti-phase with the first 
spin, disrupting the ferromagnetic coupling between the two.\\
%
%
%
For a better understanding of the correlation between the two spins, we consider Mn atoms 
 at different separations. 
 The time-dependent trajectories of the two spins, when they are nearest neighbors 
(separation $d=0.4$~nm), 
reveal that the time delay for the second spin, observed in Fig.~\ref{fig:2Mn_jx_zoom}, 
is now shorter due to closer distance. More importantly the two spins are now behaving 
as they are strongly coupled. 
Figure~\ref{fig:2Mn_ang}(a) shows the correlation function and the exponential 
fitting curves for the case when 
the Mn atoms are nearest neighbors, for 
 different values of the 
 exchange coupling (the parameters of the fitting are listed in 
 Table~\ref{tab:fitted2}). As one can see, even for very large  $J_{pd}=30$~eV,  
the spins are not completely decoupled: the decay time  
$\tau$ is roughly an order of magnitude larger than 
in the case of larger separation.  \\
Comparing 
Tables~\ref{tab:fitted} and~\ref{tab:fitted2} gives us some information about 
the dependence of the correlation between the spins on their separation. 
%
%
The characteristic decay time is consistently 
  larger at smaller separation indicating 
	 that spins stay correlated longer. Note that the strength 
	of correlation, given by parameter $a$, does not change dramatically 
	for the values of $J_{pd}$ considered here and is close to the one 
	 found at larger separation ($a$ is maximum for $J_{pd}$ and is slightly 
	 smaller than average for extremely large $J_{pd}$).\\
\begin{figure}
\begin{minipage}[h]{0.49\linewidth}
\centering
\includegraphics[scale=0.57]{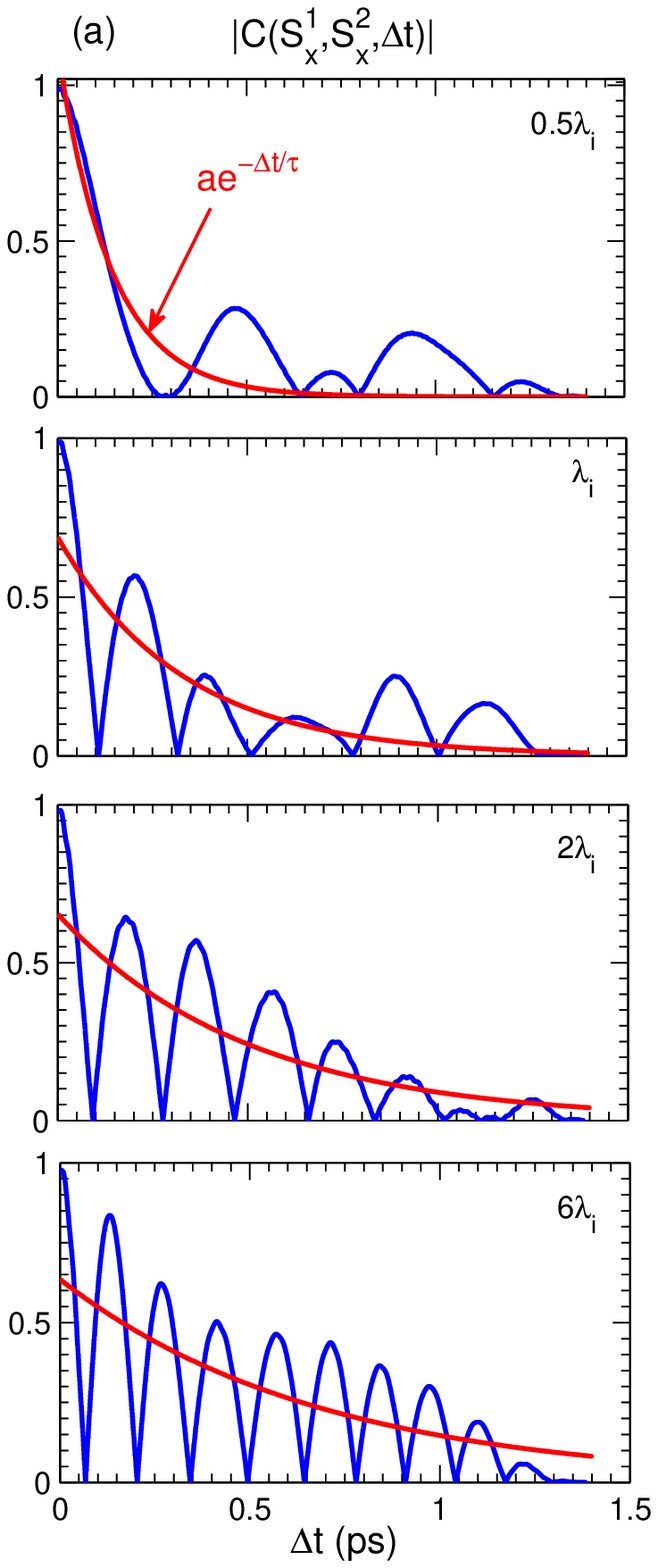}
\end{minipage}
\begin{minipage}[h]{0.49\linewidth}
\centering
\includegraphics[scale=0.57]{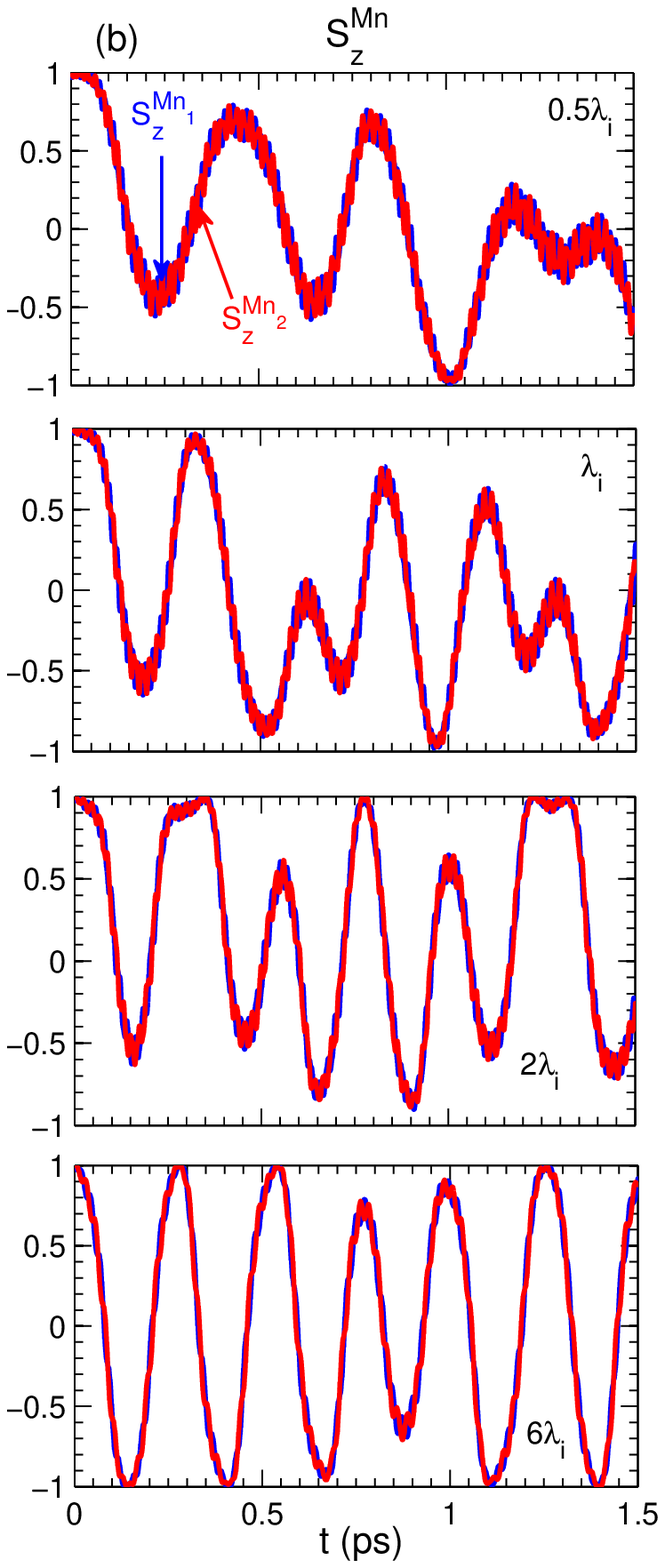}
\end{minipage}
\caption{(Color online) (a) The absolute value of 
the correlation function $\left|C(S^1_x,S^2_x,\Delta{t})\right|$, plotted as a function of the 
time lag (blue curves), and the corresponding 
 exponential fitting functions (red curves) 
 for different values of SOI. 
 (b) Time evolution of of the $z$-components of 
the two spins for the same values of SOI.}
\label{fig:2Mn_SO} 
\end{figure}
\begin{table}
\caption{\label{tab:fitted2}Parameters for the fitting  curves in Fig.~\ref{fig:2Mn_ang}
(a). Correlation function for different 
$J_{pd}$ when the Mn atoms are nearest neighbors.}
\begin{ruledtabular}
\begin{tabular}{ccc}
$J_{pd}$~(eV) & a & $\tau$~(fs) \\
\hline
0.75    &0.63    &740 \\
1.125   &0.67   &731 \\
1.5     &0.68   &719   \\
6.0     &0.85   &513   \\
30.0    &0.56   &169  \\
\end{tabular}
\end{ruledtabular}
\end{table}
\begin{table}
\caption{\label{tab:fitted3}Parameters for the fitting curves in Fig.~\ref{fig:2Mn_ang}(b). 
Correlation function for different 
initial configurations of Mn spins.}
\begin{ruledtabular}
\begin{tabular}{ccc}
the direction of the second spin at $t=0$\\ $\phi=0$& a & $\tau$~(fs) \\
\hline
$\theta=0$    &0.67    &383 \\
$\theta=\pi/18$   &0.56   &560 \\
$\theta=\pi/4$      &0.56   &452   \\
$\theta=\pi/2$      &0.45   &798   \\
$\theta=\pi$     &0.7   &22.4  \\
\end{tabular}
\end{ruledtabular}
\end{table}
\begin{table}
\caption{\label{tab:fitted4}Parameters for the fitting curves in Fig.~\ref{fig:2Mn_SO}(a). 
Correlation function for different SOI strength ($\lambda_{i}$).
The $\lambda_i$ are the reference values of the spin-orbit interaction strength for the three different atoms in the system.}
\begin{ruledtabular}
\begin{tabular}{ccc}
 $\lambda_{i}$& a & $\tau$~(fs) \\
\hline
0.5$\lambda_{i}$    &1.11    &141.6 \\
$\lambda_{i}$   &0.69   &328.2 \\
2$\lambda_{i}$     &0.65   &502.3   \\
6$\lambda_{i}$     &0.64   &682.5   \\
\end{tabular}
\end{ruledtabular}
\end{table}
It is instructive to analyze the dependence of the correlation function 
 on the initial orientation of the two Mn spins, as it 
provides insight into which configuration is more 
 favorable for long-range temporal correlations. 
 Such analysis is presented in Fig.~\ref{fig:2Mn_ang}(b). 
  At $t=0$ the first spin is pointing along $[001]$ direction 
	while the second spin is tilted by an angle $\theta$ (we set $\varphi=0$). 
	We consider a ferromagnetic (FM), $\theta=0$, antiferromagnetic (AFM), $\theta=\pi$,   
	 and non-collinear (NC), $0<\theta<\pi$, configurations. 
	The parameters for the exponential fitting 
	 functions are listed in Table~\ref{tab:fitted3}. 
 The spins are most correlated for the FM configuration. 
 In the NC case, the spins are still correlated and the 
 characteristic decay time is even larger then in the FM case. However,  
 the level of correlation is smaller, in particular for $\theta=\pi/2$.  
  For the AFM configuration, the spins become completely 
 decoupled, with the correlation function decaying rapidly within the first $\sim 20$~fs.\\
Finally, we probe the correlation function 
for different SOI [Fig.~\ref{fig:2Mn_SO}(a)]. 
 The strength of SOI directly affects the magnetic anisotropy of 
the Mn magnetic moment, i.e. the magnetic anisotropy increases with increasing $\lambda_{i}$. 
 Based on this, one might expect 
 that for larger SOI the two Mn spins will not be able  
to cross the hard plane during their time evolution (the anisotropy barrier 
in the $x$-$y$ plane defined by $\theta=\pi/2$ and arbitrary $\varphi$). 
 However, the analysis of the time-dependent longitudinal  
 components of the two spins reveals the opposite. 
According to Fig.~\ref{fig:2Mn_SO}(b), 
 $S_z^{1,2}(t)$ oscillate, switching between $[001]$ to $[00\bar{1}]$ directions,
 for all value of $\lambda_{i}$ considered. 
Therefore the anisotropy energy is not high enough to prevent  
the Mn spins from crossing the hard plane. However, as $\lambda_{i}$ increases, 
the Mn spins tend to spend less time around the hard plane and 
 oscillate more rapidly between the two easy-axis directions.\\
Interestingly, the temporal correlations between the spins become 
 stronger.  As one can see from Table~\ref{tab:fitted4}, which contains the parameters 
of the fitting functions, 
  $\tau$ increases with $\lambda_{i}$ while 
	$a$ saturates to a value of $0.6$ typical for our system. 
Similar to the case of increasing $J_{pd}$ (for $J_{pd}<10$~eV), 
 the frequency of the long-period oscillations in 
the dynamics of the Mn spins increases and becomes comparable 
 (although still smaller) to the electron ticking rate. This results 
 in more rapidly oscillating correlation functions and larger 
 decay times. \\
\section{Conclusions}
In this paper we 
studied the time evolution of substitutional Mn impurities in GaAs,
using quantum-classical 
 spin-dynamics simulations, based on the 
microscopic tight-binding model 
 and the Ehrenfest approximation. 
We described the effect of spin-orbit and exchange interactions 
 on the dynamical spectrum. A remarkable feature emerging from 
 our simulations are long-period ($\sim 5.5$~ps) oscillations of the total 
magnetic moment due to the presence of the spin-orbit interaction.\\ 
Furthermore, we studied the 
spin dynamics in a system consisting of two separated Mn 
impurities. We calculated the classical spin-spin correlation 
 functions as a measure of the effective dynamical coupling
 between the impurity spins. Our results demonstrate the  
 dependence of this coupling on the properties of the 
electronic subsystem (GaAs), as well as on the spatial 
 configuration of the impurity atoms and the initial orientation 
 of their spins.\\ 
We find that increasing the exchange 
 coupling facilitates the dynamical communication 
between the spins, mediated by the electronic degrees 
 of freedom of the host.  
However, very strong 
 exchange interaction tends to decouple the two spins 
as the host carriers become localized.  
 For smaller spacial separation between the impurity 
 atoms, the characteristic decay time of the correlation 
function increases significantly for all values of $J_{pd}$. 
  When the spins are initialized 
	  in the same direction (or slightly tilted) they remain ferromagnetically 
		coupled, with a characteristic in-phase oscillation pattern,  
	over times of the order of few hundred femtoseconds. 
 In contrast to this, when starting from an antiferromagnetic 
configuration, the decay time decrease to tens of femtoseconds. 
 Increasing the spin-orbit interaction strength 
 seems to have a similar effect as the exchange interaction 
 (at least for small values of $J_{pd}$), 
 i.e. the temporal correlations between the two spins become 
more long-ranged. We note, however, that the energy scale 
 associated with the spin-orbit interaction is orders of magnitude smaller than 
that of the exchange interaction.\\
The role of the electronic subsystem of the host material, as well as 
the effects of inter- and intra-atomic interactions 
 (exchange and spin-orbit interactions), 
on the dynamics of individual atomic-scale magnets 
are important questions that are becoming 
accessible experimentally. The present study contributes to fundamental 
understanding of spin dynamics in such devices, which is 
 crucial for future applications in solotronics.\\ 
%
%
\begin{acknowledgments}
This work was 
supported by the Faculty of Technology
at Linnaeus University, by the
Swedish Research Council under Grant Number: 621-2010-3761, 
and the NordForsk research network 080134 ``Nanospintronics: theory and
simulations".
Computational resources have
been provided by the Lunarc center for scientific and technical computing at
Lund University.
\end{acknowledgments}
%
%
\bibliography{sd}
%
\end{document}